\documentclass[twocolumn]{aa}
\usepackage{graphicx}
\usepackage{epsfig}
\begin{document}
\newcommand{\gsim}{\hbox{\rlap{$^>$}$_\sim$}}                               
\titlerunning{On the origin Of X-Ray Flashes }
\title{On the origin of X-ray Flashes }

\author{Shlomo Dado\inst{1} \and Arnon Dar\inst{1,2} \and
A. De R\'ujula\inst{2}}
\institute{Physics Department and Space Research Institute, Technion,
               Haifa 32000, Israel\\
           \and Theory Division, CERN, CH-1211 Geneva 23, Switzerland}


\maketitle

\abstract

{{\bf ABstract.} We use the cannonball (CB) model of gamma ray
bursts (GRBs) and their afterglows (AGs) to analyze the observational data
on X-ray flashes (XRFs) and their AGs.  We show that the observations
support the CB-model interpretation that XRFs, like GRBs, are produced by
the explosions of core-collapse supernovae (SNe) akin to SN1998bw, by jets
of highly-relativistic CBs. The XRFs and GRBs are intrinsically identical
objects, but the XRFs are viewed from angles (relative to the jet
direction) which are typically a few times larger than the typical viewing
angles of ``classical'', long-duration GRBs.  There should be
XRFs, not observed so far, with durations similar to those of short GRBs.

}

\maketitle

\section{Introduction and summary}

By definition, XRFs are GRB-like bursts of photons whose ``peak energy'',
$E_p$, is below 40 keV (roughly speaking, $E_p$ is the maximum of the
distribution $\nu\,F_\nu=E^2\,dN_\gamma/dE$).  They are rich in X-rays but
relatively poor in $\gamma$-rays, as implied by their name.  They were
discovered with the Wide-Field Camera of the Beppo-SAX satellite, but they
were not seen above 40 keV with its GRB Monitor (Heise et al.~2001). They
were detected by BeppoSAX at a rate of 4 per year, indicating a population
not very much smaller than that of GRBs\footnote{GRBs were detected by
BATSE at a rate of $\sim 1$ per day, but with much higher sky-coverage
than BeppoSAX has.}. Re-examining the BATSE data, Kippen et al.~(2002)
have found some 10 XRFs. A few more have been detected by HETE II. To
date, about 30 XRFs have been reported. These bursts are distinguished
from Galactic transient sources by their isotropic spatial distribution,
analogous to that of GRBs.  They are softer and dimmer than GRBs, but
otherwise their properties are similar: they have similar durations and
light curves, their spectrum is also well described by the ``Band
spectrum" (Band et al.~1993) and their spectral evolution is similar to
that of GRBs.

To date, four XRFs (011030: Gandolfi et al.~2001; 020427: in 't Zand et
al.~2002; 020903: Ricker et al.~2002;  030723: Prigozhin et al.~2003) have
been followed up sufficiently rapidly to allow for the detection of their
afterglows (AGs), i.e.~continued lower-energy emissions, and for their
precise (sub-arcsecond)  localization. The AGs of XRFs were first
discovered by Taylor et al.~(2001) in the radio band, by Harrison et
al.~(2001) in the X-ray band, and by Soderberg et al.~(2002) in the
optical band. A host galaxy of an XRF was discovered first by Fruchter et
al.~(2002). The early-time AGs of XRFs have
spectral and temporal properties similar to those of the AGs of
the ``classical'' long-duration GRBs, but they are dimmer.  
In three cases the precise localization afforded by the AG observations led to an
identification of the putative host galaxies of the XRFs progenitors
(Soderberg et al.~2002;
Bloom et al. 2003a). The few available redshift and photometric
informations on these host loci indicate that XRFs are cosmological in origin, and
that their progenitors, like those of long-duration GRBs, 
permeate star-formation regions of their host galaxies.
Not all measured XRF redshifts are high enough to explain their relatively low
peak flux and low peak energy simply on grounds of distance and cosmological
redshift.

In the CB model (Dar \& De R\'ujula 2000, 2003; Dado, Dar \& De R\'ujula 
2002a, 2003a and references therein) {\it long-duration} GRBs and 
their AGs are produced 
in  {\it ordinary core-collapse} SN explosions
by jets of CBs, made 
of {\it  ordinary atomic matter}. Each CB gives rise
to a single $\gamma$-ray pulse of a GRB's light curve. The CBs
are initially travelling with high Lorentz factors,
$\gamma$, which peak 
very narrowly around
an initial $\gamma\sim 10^3$ (Dado et al. 2002a, 2003a). The light CBs
emit is collimated in a cone of opening angle $\sim 1/\gamma\sim 1$ mrad.
The ``jet opening angle'' $\theta_v$ (subtended by a CBs' transverse
radius as viewed from their point of emission) is smaller than $1/\gamma$ 
and its effects can be neglected: in this respect, it is as if
CBs were point-like. The differences between the pulses of 
different GRBs are dominated
by the different values of $\theta$, the observer's viewing angle relative
to the CB's direction of motion (Dar \& De R\'ujula 2000, 2003).

The properties of XRFs are similar to those of GRB 980425, which, in the
CB model, was interpreted (Dar \& De R\'ujula 2000, 2003;  Dado et
al.~2002a, 2003a) as an entirely normal GRB produced by the
explosion of SN1998bw (Galama et al.~1998). Its jet of CBs was ejected at
an angle $\theta \sim 3.9/\gamma\sim 8$ mrad, a large value which,
combined with the progenitor's unusually small redshift ($z=0.0085$)
conspired to produce a rather typical GRB fluence (Dar \& De R\'ujula
2000, 2003; Dado et al.~2002a, 2003a). GRB 980425 is by
definition a GRB and not an XRF, as the central value of its peak energy,
$E_p=54.6\pm 20.9$ keV (Yamazaki, Yonetoku \& Nakamura 2003a), 
is just above the ``official'' borderline 40 keV.
In the CB model, SN1998bw, associated with GRB 980425, is an 
ordinary core-collapse SN: its ``peculiar'' X-ray and radio emissions 
were not emitted by the SN, but were part of the GRB's AG (Dado 
et al. 2002a, 2003a). The high velocity of its ejecta is attributed
to the SN being viewed almost ``on axis"

The  CB-model interpretation
of XRFs (Dar \& De R\'ujula 2003) is entirely straightforward: {\it XRFs
are GRBs viewed further off axis}\footnote{Subsequent to
critiques (Dar 1998, 1999, 2003;  Dar \& Plaga 1999; Dado et al.~2002a; De
R\'ujula 2002, 2003) of the 
{\it on-axis} viewing-angle anzatz, $\theta=0$, 
of the ``collimated-fireball'' or conical-jet models (e.g.~Rhoads
1997, 1999; Sari, Piran \& Halpern 1999; Piran 2000; Panaitescu and Kumar
2000; Frail et al.~2001; Berger et al.~2003; Bloom et al.~2003b),
 this  assumption has
recently been replaced by a $\theta\neq 0$ of uniform or 
structured conical jets, e.g.~Rossi, Lazzati \& Rees 2002; Granot et
al.~2002; Zhang \& M\'esz\'aros 2002; Panaitescu \& Kumar 2003;  Jin \&
Wei 2003; Salmonson 2003. The jets' opening angles are $\theta_v>1/\gamma$,
but getting  close to inverting the inequality (Waxman 2003,
Lazzatti et al.~2003) to spouse the CB-model's geometry.}.
Thus, all of the CB-model results (e.g.~Dado et al.~2002a, 2003a)
are applicable to XRFs. 

In this paper we show that the CB model explains well all the observed
properties of XRFs and their AGs.  In particular, their large
viewing angles result in much smaller Doppler factors, $\delta$,
than those of GRBs. For the Lorentz factors and viewing angles
relevant to our discussion,
$\gamma^2 \gg 1$ and $\theta^2 \ll 1$, and
\begin{equation}
  \delta = {1\over \gamma\, (1-\beta\, \cos\theta)}\approx
           {2\gamma\over 1+\gamma^2\, \theta^2}\; ,
\label{Doppler}
\end{equation}
to an excellent approximation. 

Relative to GRBs, XRFs have pulses that are dimmer in fluence
by a factor $\delta^3$ (or in photon number-count by a factor $\delta^2$).
The XRF pulses are wider and their peak energies are smaller than
those of GRBs, both by a factor $\delta$. Their 
AGs are dimmer and less rapidly-evolving at early times, in a 
manner that we shall describe. On the other hand, both
for XRFs and GRBs, the total duration of multi-pulse events is
governed by the properties of the engine generating the CBs: the observed
duration and the local duration at the progenitor's position are related
by a cosmological time-stretching factor, $1+z$, which does not involve $\delta$.
The total duration of XRFs is therefore akin to that of GRBs, while their
structure is ---since they consist of similar numbers of wider pulses---
smoother than that of GRBs. Finally, in the AGs of relatively nearby XRFs it 
should be possible to observe the contribution of a SN akin 
to SN1998bw, displaced to the XRF's position and peaking about one month
after the SN exploded and the XRF was emitted
(Dar \& De R\'ujula 2003). Such a shining-gun
signature may already have been observed in the AG of XRF
030723 (Fynbo et al.~2003), see Fig.~(\ref{figXRF}).

In the CB model, X-ray rich GRBs fall naturally in an intermediate
domain between GRBs and XRFs: they do not require a separate discussion.

The interpretation of XRFs as GRBs viewed off axis has also been advocated
by other authors (e.g. Yamazaki, Ioka \& Nakamura 2002, 2003b; Jin \& Wei,
2003).  One main difference between our CB-model and theirs is the jet
geometry: effectively point-like CBs, versus conical shells. Another one
is the underlying GRB and XRF production mechanism: inverse Compton
scattering by the CBs' electrons of the ``ambient light'' that permeates
the wind-fed surroundings of the parent star, versus synchrotron radiation
emitted in collisions of conical shells of a delicately baryon-loaded
$e^+e^-$ plasma, produced by ``collapsars'' (Woosley \& MacFadyen 1999) or
``hypernovae'' (Iwamoto et al.~1998). Finally, the AGs are in both models
generated by synchrotron radiation, but in the CB model the process does
not involve shocks (Dar \& De R\'ujula 2003) and results in a description
of AG light-curves and wide-band spectra  simpler and more successful than 
that of the fireball or firecone models (Dado et al. 2002a, 2003a and, 
for a direct comparison, 2003d).

\section{The engine}

In the CB model, long-duration GRBs  are produced 
in the explosions of ordinary core-collapse SNe.
Following the collapse of the stellar core into a neutron
star or a black hole, and given the characteristically large
specific angular momentum of stars, an
accretion disk or torus is hypothesized to be produced around
the newly formed compact object, either by stellar material originally   
close to the surface of the imploding core and left behind by the
explosion-generating outgoing shock, or by more distant stellar matter
falling back after its passage (De R\'ujula 1987). A CB is emitted, as
observed in microquasars, when part of the accretion disk
falls abruptly onto the compact object  (e.g.~Mirabel \& Rodrigez 1999;
Rodriguez \& Mirabel 1999 and references therein).
The high-energy photons of a single pulse in a GRB or
 an XRF are produced as a CB coasts
through the ``ambient light'' permeating the surroundings of the
parent SN. The electrons enclosed in the CB Compton
up-scatter photons to energies that, close to the CBs direction
of motion, correspond to the $\gamma$-rays of a GRB and
less close to it, to the X-rays of an XRF.
Each pulse of a GRB or an XRF corresponds to one
CB. The timing sequence of emission of the successive individual pulses
(or CBs) reflects the chaotic accretion process and its
properties are not predictable, but those of the single pulses are (Dar \&
De R\'ujula 2003 and references therein).

The observational evidence prior to GRB 030329 for the claim that
core-collapse SNe are the engines generating GRBs, (e.g. Dar \& Plaga
1999; Dar \& De R\'ujula 2000), is discussed in Dado et al.~2002a;
2003c. The spectroscopic discovery of SN2003dh in the AG of GRB 030329
(Garnavich et al.~2003; Stanek et al.~2003; Hjorth et al. 2003) with a
luminosity and spectrum remarkably similar to those of SN1998bw, as
predicted by Dado et al.~(2003c), has provided the most convincing
evidence for the GRB-SN association after GRB980425/SN1998bw.

\section{Properties of the X-rays of XRFs}

Since, in the CB model, XRFs are GRBs viewed from a larger
angle $\theta$, all we have to do to study or predict the properties
of the former is to paraphrase, for larger $\theta$, the properties of the 
latter, for which the CB model offers a general,
simple and predictive description (Dar \& De R\'ujula 2003).

If core collapse SNe and their environments were all identical, and if
their ejected CBs were also universal in number, mass, Lorentz factor and
velocity of expansion, all GRBs would be standard candles and
the observed differences between them would only be due to
the observer's position, determined by $z$ and the angle of observation,
$\theta$. The distribution of Lorentz factors inferred within
the CB model can be inferred both from the study of GRB AGs
(Dado et al. 2002a, 2003a), and from that of the GRBs themselves (Dar
\& De R\'ujula 2003). This distribution is very
narrowly peaked around $\gamma\simeq 10^3$, 
so that the dependence of the various observables
on the wider distribution of $\theta$-values directly 
translates into their dependence on $\delta(\theta,\gamma)$
at an approximately constant $\theta$. This dependence
is strong in various observables, such that it might overwhelm much of the
case-by-case variability induced by the distributions of the other
parameters.

\subsection{The viewing angle of XRFs}
The mean peak energy  of ordinary GRBs is
$\langle E_p(GRB)\rangle\approx 250$ keV (e.g. Preece et al.~2000; Amati et 
al.~2002).
It is convenient to introduce the quantity:
\begin{equation}
\sigma\equiv{\gamma\,\delta\over 10^6}\,{2\over 1+z}\; , 
\label{sigma}
\end{equation}
to which the peak energy of GRBs and XRFs is proportional,
since $\gamma\delta/(1+z)$ is the factor by which, on average, 
electrons with a Lorentz factor $\gamma$ boost the energy
of a ``target photon'' of an isotropic distribution into a
final photon observed at an angle $\theta$ (Dar \& De R\'ujula 2003).
The mean $\sigma(GRB)$ of GRBs of known $z$ is
$\langle\sigma(GRB)\rangle\approx 1$. 

The ratio between the peak
energy $E_p(XRF)$ of an XRF and the mean peak energy
of GRBs is:
\begin{equation}
E_p(XRF)={\sigma(XRF)\over\langle\sigma(GRB)\rangle}
\; \langle E_p(GRB)\rangle \, .
\label{Epeak}
\end{equation}
The defining condition of an XRF, $E_p(XRF)<40$ keV,
implies that $\sigma(XRF)$ must be smaller by a factor
$\sim 6.25$ than $\langle\sigma(GRB)\rangle\approx 1$.
The mean redshift of GRBs, as inferred from the current sample of  
32 GRBs with known redshift, is $\langle z\rangle\approx 1\, .$
By use of Eq.~(\ref{Doppler}), and adopting the characteristic value  
$\gamma\sim 10^3$, we learn that 
the typical viewing angles of XRFs satisfy 
$\theta_{XRF} \geq 3.4$ mrad.
For GRBs the result is 
$\langle\theta\rangle_{GRB}\sim 1$ mrad (Dar \& De R\'ujula 2003).

The dependence of many other GRB observables on the viewing angle or the
Doppler factor (Dar \& De R\'ujula 2000, 2003; Plaga 2001) has direct
testable consequences.

\subsection{Polarization}

 Inverse Compton scattering of the ambient light by the
CBs' electrons linearly polarizes it. The degree of polarization
$\Pi(\theta,\gamma)$ is a function of only the product $\theta\,\gamma$
and has the universal shape: (Shaviv \& Dar 1995; Dar \& De R\'ujula
2003):
\begin{equation}
\Pi(\theta,\gamma)\approx {2\;\theta^2\,\gamma^2\over 
1+\theta^4\,\gamma^4}\, .
\label{polSN}
\end{equation}  
For XRFs $\theta\, \gamma \geq 3.4$ and Eq.~(\ref{polSN})
yields a polarization level, $\Pi \leq 17\%$. 
Thus, the CB model predicts a much smaller  linear polarization
in XRFs than in GRBs, which reach $\Pi \approx  100\%$
for $\theta\approx 1/\gamma$.   

\subsection{Duration of single pulses of XRFs} 

The times, $t'$ in the CB's rest  frame,
$t_{_{SN}}$ in the SN rest frame, and $t$ as measured by a distant
observer, are related via $ dt=(1+z)\, dt'/\delta=(1+z)\, dt_{_{SN}}/
\gamma\, \delta$. Hence, any time measure $\Delta t$ in a single 
pulse in
a XRF, such as its full width at half maximum, its rise time, its fall
time, or its ``lag time'' (the difference
between the pulse peak times in two different energy intervals),
is proportional to $(1+z)/\delta$. In a ``standard 
candle'' approximation, it can be written as:
\begin{equation}
\Delta t (XRF)={\langle \Delta t\, E_p\rangle_{_{GRB}}
\over E_p(XRF)}\approx 
6.25\; {40\, {\rm keV}\over E_p(XRF)}\, \langle \Delta t\, 
\rangle_{_{GRB}}\; .  
\label{duration}
\end{equation} 
The  durations of single pulses in
XRFs, for which $\theta \geq 3.4/\gamma$, should be, on average, longer 
than  in GRBs by a factor $\geq 6.25$. 
The median duration of the single pulses of GRBs is 
$\overline{\Delta t}(GRB)\sim 0.92$ s 
full width at half-maximum (Lee, Bloom \& Petrosian 2000; 
McBreen et al.~2002). The prediction for
XRFs is, therefore, $\overline{\Delta t}(XRF)\geq 5.75$ s.

\subsection{The total duration of XRFs} 

The duration of a multi-pulse GRB or XRF is the total duration 
of the CBs' emission-times by the CB-emitting engine, 
which is independent of the viewing angle, plus the duration
of the last pulse. On average, there
are 6 significant pulses in a single GRB (Quilligan et al.~2002)
and the interval between pulses is, on average, roughly
twice a pulse's duration. Thus, it is a fair approximation 
(on average) to neglect a pulse's duration relative to the total duration
of a GRB or an XRF. In that approximation, their total
durations should be approximately the same. This agrees with
the observations
(Heise et al. 2000; in 't Zand et al. 2002; Kippen et al. 2002).

In contrast with the above, the duration of individual pulses is,
on average, $\sim 6.25$ times
longer in XRFs than in GRBs, as discussed in the previous
subsection. The $1+z$ redshift dependence of the total-
and pulse durations is the same, and it is common to
XRFs and GRBs. Hence, XRFs should have a much less 
pronounced temporal structure and a much smaller
``variability'' than GRBs.

\subsection{XRF Fluences} 

The total energy emitted by a CB in its rest system,
${\cal{E}}'_{_{CB}}\sim 0.8\times 10^{44}$ erg on average,
is found to be very narrowly peaked around its central
value\footnote{In the CB model GRBs (and, more so, their individual
pulses) are much better standard candles than
in the FB models (Frail et al.~2001; Berger et al.~2003; Bloom et al.
2003b).}: its dispersion is a factor $\sim 2$ either way
(Dado et al.~2002a, Dar \& De R\'ujula 2003). Let
$D_L(z)$ be the luminosity distance (7.12 Gpc at $z=1$, for the
current cosmology with $\Omega=1$, $\Omega_\Lambda=0.7$ and $H_0=65$
km s$^{-1}$ Mpc$^{-1}$). Let 
$G(\theta)\approx 2\, (1+\theta^4\,\gamma^4)/ (1+\theta^2\,\gamma^2)^2$  
be  the angular dependence of the low-energy Klein-Nishina
cross section for Compton scattering, corresponding ---in the
large $\gamma$, small $\theta$ limit--- to
$d\sigma/d\cos\theta'\propto 1+\cos^2\theta'$ in a CB's rest system.

In the fairly good standard-candle approximation, the observed
energy fluence of a single CB is given by:
\begin{equation}
F_{_{CB}}
 = {(1+z)\, G(\theta)\, \delta^3
                  \over 4\, \pi\,  D_L^2}\; {\cal{E}}'_{_{CB}}\, .
\label{Fluence}
\end{equation}  
To a good approximation, $G=2$ for  $\theta\, \gamma\geq 3.4\, .$
Because of the strong $\delta$-dependence, the energy fluence of XRFs 
is a factor $\geq  6.25^3/2\approx 122$ smaller than that of GRBs,
for which $G(\theta)\sim 1$.

In a sample of 35 XRFs and GRBs well localized  with  HETE II (Barraud et 
al.~2003), the mean energy fluence in the 7--400 keV band  of 19 GRBs
whose peak energy was well constrained by the observations,  
was $264\times 10^{-7}\,  \rm {erg\, cm^{-2}}$. This is to be compared with 
$\sim 3\times 10^{-7}\,  \rm {erg\, cm^{-2}}$ for the 7 XRFs in the 
sample. The corresponding GRB--to--XRF fluence ratio is  $\sim 83$,
in satisfactory agreement with the expectation, particularly in view  
of the fact that the truncation of the contribution from  
$E_\gamma \geq 400$ keV  subtracts a larger fraction from
the GRBs' fluence than from the XRFs' fluence.   

\subsection{Correlations}

The $\delta$-dependence is also strong in observables
other than the fluence. To give a few examples, $\Delta t\propto 1/\delta$;
$E_p\propto\delta$;
the photon-number fluence, $f$, is 
$\propto\delta^2$; the peak photon flux,  $f_p$, and the ``isotropic'' energy 
of a pulse, ${\cal{E}}^{iso}$,  
are  $\propto\delta^3$; and the
peak luminosity $L_p$ (energy fluence per unit time)
is  $\propto\delta^4$ (Dar \& De R\'ujula 2000, 2003).
The powers of $\delta$ involved in these proportionality factors imply, 
among others, the following correlations:
\begin{equation}
E_p\propto [f]^{1/2} \, ;~~~~ \Delta t\propto [f_p]^{-1/2}\, ,
\label{epgflu}
\end{equation}
\begin{equation}
E_p\propto [f_p]^{1/3} \, ;~~~~ \Delta t\propto [f_p]^{-1/3}\, ,
\label{epint}
\end{equation}
\begin{equation}
E_p\propto F^{1/3}\, ; ~~~~~ \Delta t\propto [F]^{-1/3}\, ,
\label{epflu}  
\end{equation}
\begin{equation}
E_p\propto [{\cal{E}}^{iso}]^{1/3}\, ; ~~~~
\Delta t\propto[{\cal{E}}^{iso}]^{-1/3}\, ,
\label{epeiso}
\end{equation}
\begin{equation}
E_p\propto [L_p]^{1/4}\, ;~~~~ \Delta t\propto [L_p]^{-1/4}\, .
\label{epilum}
\end{equation}

The correlations in Eqs.~(\ref{epgflu}) to (\ref{epilum}) apply
to individual pulses and pulse averages over a GRB or an XRF.
The relations not involving time measures $\Delta t$
(which behave differently for pulses and inter-pulse delays)
also apply to global XRF or GRB properties. All these correlations, 
are well satisfied for ordinary GRBs in particular for pulses of GRBs with known
redshift, after correction for the $z$-dependence
(Dar \& De R\'ujula 2003). At present, the low photon
counting-levels in XRFs prevent a reliable resolution of XRFs
into individual pulses, and the ensuing tests of most of the above
correlations for single pulses.

\subsection{The shape of single pulses} 
In the CB model, to a rather good approximation, single GRB and XRF
pulses have shapes that can be approximated by the simple expression:
\begin{equation}
{dN\over dt}=exp\left[-\left({t_1\over t}\right)^m\right]\,
\left\{1-exp\left[-\left({t_2\over t}\right)^n\right]\right\}\, ,
\label{pheno}
\end{equation}
where $t$ is the time from the beginning of the pulse (not of the GRB
or XRF),
$t_1$ and  $t_2$ are the characteristic rise and fall times,
($t_i={\cal{O}}(\Delta t)$ in Eq.~(\ref{duration}) for XRFs),
and $n$ and $m$ are power indices whose median 
values are near  2. This pulse shape describes well the observed
shape of single GRB pulses (Dar \& De R\'ujula 2003).
At present, low statistics prevents a reliable resolution of XRFs 
into individual pulses and a test of this prediction.

\subsection{Spectral shape and spectral evolution} 

In the CB model, the $\gamma$ rays of a GRB and the 
X-rays of an XRF are generated by inverse Compton scattering
of ``ambient light''. This light permeates the semi-transparent,
previously wind-fed surroundings of the parent star and results
from their illumination by the early SN luminosity. The ambient
light, as befits a semi-transparent, externally-energized medium,
has a thin thermal-bremsstrahlung spectrum 
$dN/dE_i\propto exp(-E/T_i)/E$, with $T_i={\cal{O}}(1)$ eV.
The ambient light is Compton up-scattered by the CBs' electrons,
some of which are simply comoving with it, while others
have been accelerated and Compton-cooled to a power-law 
distribution of index $\tilde p\sim 3.2$, by a ``shockless
acceleration'' process studied in Frederiksen et al.~(2003) 
and discussed in Dar \& De R\'ujula (2003). The result of the convolution
---with an inverse Compton scattering kernel---
of the electron and ambient light distributions is, to a 
very good approximation, given by the simple expression:
\begin{eqnarray}
{dN\over dE}
&\propto&
\left({T\over E}\right)^\alpha\; e^{-E/T}+b\;
(1-e^{-E/T})\;
{\left(T\over E\right)}^\beta\nonumber\\
\alpha&\sim&1\; ; \;\;\;\;\;\;\;\;
\beta={{\tilde{p}}+1\over 2}\sim 2.1\, .
\label{totdist}
\end{eqnarray}
Here $b$ is a parameter reflecting the different normalizations
of the comoving and accelerated electrons and
\begin{equation}
T\equiv {4\over 3} \;T_i\;{\gamma\;\delta\over 1+z}\;\,\langle 1+\cos\theta_i\rangle,
\label{Teff}
\end{equation}
where $\langle\cos\theta_i\rangle$ is the mean angle of the ambient light
photons relative to the radial direction towards the progenitor, reflecting
the fact that, for a semitransparent wind, the ambient light may not
be perfectly isotropic.

\begin{figure}[t]
\hskip 2truecm
\vskip -3.5cm
\vspace*{3.8cm}
\epsfig{file=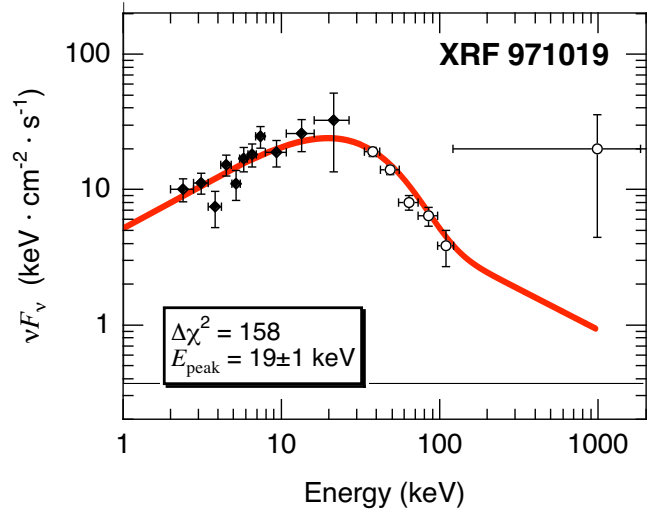, width=8.5cm} \\
\vspace{-0.5cm}
\caption{The spectral data from BeppoSAX/WFC (solid diamonds) and
CGRO/BATSE (open circles) on GRB 971019 (Kippen et al.~2002)
and the CB-model fit of Eq.~(\ref{totdist}) to these data. The $\chi^2$
and $E_p$ values are those of the Band fit. The CB-model fit has
$T\sim E_p$ and better $\chi^2$.}
\label{XRF971019}
\end{figure}

\begin{figure}[t]
\vspace{-3.0cm}
\hskip 2truecm
\hspace*{-2.8cm}
\epsfig{file=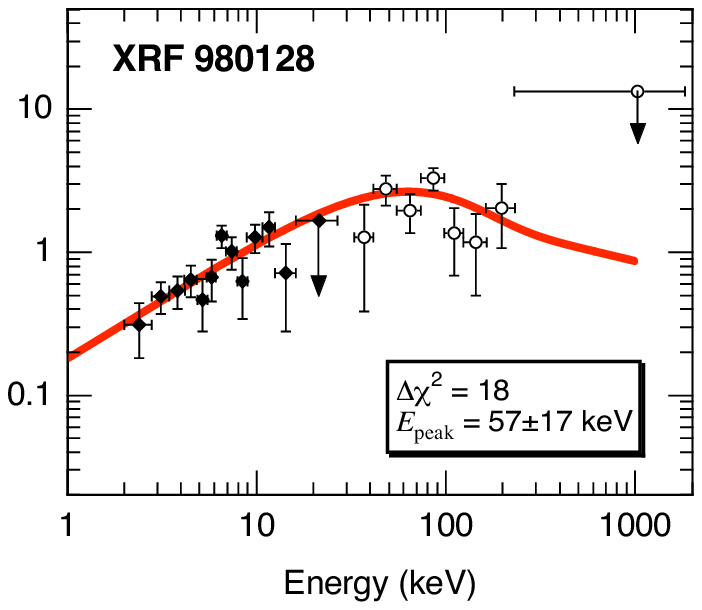, width=23cm} \\
\vspace{-22.5cm}
\caption{The spectral data from BeppoSAX/WFC (solid diamonds) and
CGRO/BATSE (open circles) on GRB 9810128 (Kippen et al.~2002)
and the CB model fit of Eq.~(\ref{totdist}) to these data. 
The $\chi^2$
and $E_p$ values are those of the Band fit. The CB-model fit has
$T\sim E_p$ and similar $\chi^2$.}
\label{XRF980128}
\end{figure}

\begin{figure}[t]
\vspace{-3.5cm}
\hskip 2truecm
\hspace*{-3.1cm}
\epsfig{file=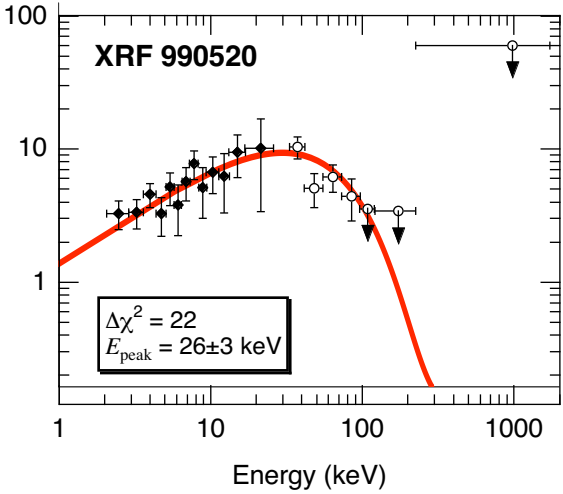, width=22cm} \\
\vspace{-21cm}
\caption{The spectral data from BeppoSAX/WFC (solid diamonds) and
CGRO/BATSE (open circles) on GRB 990520 (Kippen et al. 2002)
and the CB model fit of Eq.~(\ref{totdist}) to these data. The $\chi^2$
and $E_p$ values are those of the Band fit. The CB-model fit has
$T\sim E_p$ and similar $\chi^2$.}
\label{XRF990520}
\end{figure}

The spectrum of Eq.~(\ref{totdist}) is independent of the CB's expansion
rate, its baryon number, its geometry and its density profile. Moreover, 
its derivation rested only on observations of the properties of the
surroundings of exploding stars, Coulomb scattering, and an
input electron-distribution extracted from numerical studies also
based only on ``first principles''.
The spectrum predicted in the CB model bears a striking
resemblance to the phenomenological Band spectrum traditionally used 
to describe GRB energy spectra (e.g.~Band et al.~1993: Preece et al.~2000 
for an analysis of BATSE data, Amati et al.~2002 for BeppoSAX data,
and Barraud  et al.~2003 for HETE II data).
This spectrum also fits well the spectrum observed in XRFs as shown 
in Figs.~(\ref{XRF971019}), (\ref{XRF980128}) and (\ref{XRF990520}).
The parameters of the fits are, respectively,
$T=20$, 50, 30 keV, $b=0.2$, 0.7, $\sim 0.01$, and $\beta=2.6$, 2.3, 2.7.
The fits, in particular the last one, are quite insensitive to $b$.


\subsection{The time--energy correlations}

Such as we have treated it so far, the distribution $dN/dt\,dE$
of the X-rays in an XRF pulse ---as a function of both time and
energy--- is a product of a function of only time, Eq.~(\ref{pheno}), and a 
function of only energy, Eq.~(\ref{totdist}). One reason for this
is that we have
not yet taken into account the fact that the cooling time
of the accelerated electrons in a CB ---by Compton scattering--- is of the 
same order of magnitude as the (Compton-scattering) transparency time 
of the CB, which is the time characterizing a pulse's width
(Dar \& De R\'ujula 2003).
Consequently, the index $\tilde{p}$ of the power-law
electron energy distribution ought to evolve during a pulse
from $\tilde{p}\sim 2.2$ to
$\tilde{p}\sim 3.2$. Equivalently, the index $\beta$ in 
Eq.~(\ref{totdist}) is expected to vary from $\beta=\beta_1\sim 1.6$ to 
$\beta=\beta_2\sim 2.1$.

There is another fact contributing to a non-trivial
correlation between energy and time within a GRB pulse.
The relation between the ``temperature'' $T_i$ characterizing
the initial ambient-light thermal-bremsstrahlung distribution
and that of the resulting GRB or XRF photons,
$T$, is that of Eq.~(\ref{Teff}). 
As the CB reaches the more transparent
outskirts of the wind, its ambient light distribution is bound to become
increasingly radial, meaning that the average $1+\cos\theta_i$ in 
Eq.~(\ref{Teff}) will depart from $\sim 1$ and tend 
to $0$ as $1/r^2$: the point-source long-distance limit. Since this transition
depends on the integrated absorption by a wind with 
$\rho\propto 1/r^2\propto 1/t^2$,
it can be characterized by a simple time-dependence
of the effective temperature in Eq.~(\ref{totdist}):  
\begin{equation}
T\to T(t)\sim T\,\{1-exp[-(t_T/t)^2]\},
\label{Tevol}
\end{equation}
with $t_T$ of the same order as a pulse's width, or is
somewhat smaller (Dar \& De R\'ujula 2003).

The two time-energy correlations we just discussed make pulses
decrease in duration and peak earlier in time as the energy interval
in which they are measured increases. They also imply that a
pulse's spectrum gets softer as time elapses during a pulse.
This behaviour is observed in GRBs (e.g.~Fenimore et al.~1995; 
Norris et al.~1996; Ramirez-Ruiz \& Fenimore 2000; Wu \& Fenimore 2000;
Golenetskii et al.~1983;  Bhat et al.~1994). We expect XRFs to
display precisely the same trends.
The spectral evolution in a GRB and XRF pulse is mainly due to the decline of 
the effective temperature described by Eq.~(\ref{Tevol}).
This implies that the width of a GRB on an XRF
pulse, dominated by its late-time behaviour, 
changes as it is measured in different energy bands approximately as:
\begin{equation}
\Delta t \propto E^{-\kappa},\;\;\;\;\;\;\; \kappa\; \leq \;0.5,
\label{fenimore}
\end{equation}
where $\kappa=0.5$ is the limiting value for an exact $T\propto 1/t^2$.
This result is in agreement with the observationally inferred  relation
$t_{_{\rm FWHM}}\propto E^{-0.43\pm 0.10 }$
for the average full width at half-maximum
 of GRB pulses  as a function of the energies
of the four BATSE channels (Fenimore et al.~1995, Norris et al.~1996).
We expect this result to be valid for XRF pulses as well.

\section{The afterglow of XRFs}

In the CB model, the AGs of GRBs and XRFs consist of three contributions,
from the CBs themselves, the concomitant SN, and the host galaxy:
\begin{equation}
F_{AG}=F_{CBs}+F_{SN}+F_{HG}\, ,
\label{sum}
\end{equation}
the latter contribution being usually determined
by late-time observations, when the CB and SN contributions become
negligible. 

Let the unattenuated energy flux
density of SN1998bw at redshift $z_{bw}=0.0085$ (Galama et al. 1998)   
be $F_{bw}[\nu,t]$. For a similar SN placed at a redshift $z$
(Dar 1999; Dar \& De R\'ujula 2000):   
\begin{equation}
F_{SN}[\nu,t] = {1+z \over 1+z_{bw}}\;
{D_L^2(z_{bw})\over D_L^2(z)}\, A(\nu,z)\, F_{bw}[\nu',t']\, ,
\label{bw}
\end{equation}
where $\rm D_L(z)$ is the luminosity distance, 
$A(\nu,z)$ is the attenuation along the line of sight,
$\nu'=\nu\, (1+z)/ (1+z_{bw}),$  and $t'=t\, (1+z_{bw})/(1+z)$.     

The evolution of
$F_{_{CBs}}$ with time is due to the deceleration of the CBs in the ISM.
In an approximately hydrogenic ISM of constant number density
$n_p$, the function $\gamma(t)$ is determined by energy-momentum
conservation to be the real root of the cubic:
\begin{equation}
{1\over\gamma^3}-\rm{1\over\gamma_0^3}
+3\,\theta^2\,\left[{1\over\gamma}-{1\over\gamma_0}\right]=
{2\,c\, t\over 3\, (1+z)\, x_\infty}\; ,
\label{cubic}
\end{equation}
where $\gamma_0$ ---previously called simply $\gamma$---
is the initial Lorentz factor of a CB (very close to being
constant during the GRB or XRF emission),
$t$ is the observer's time and
\begin{equation}
x_\infty\equiv{N_{_{B}}\over\pi\, R_\infty^2\, n_p}\, ,
\end{equation}
with $N_{_{B}}$ being the baryon number of the CB and  $R_\infty$ the 
calculable asymptotic radius of a CB, reached
within minutes of observer's time (Dado et al.~2002a). 
It takes a distance $x_\infty/\gamma_0$, typically a fraction of a kpc, for
the CB to half its original Lorentz factor.

The AG of the CBs is mainly due to synchrotron radiation from 
accelerated  electrons in the CB's
chaotic magnetic field. It has the approximate form 
(Dado et al.~2003e):
\begin{equation}
F_{_{CB}}[\nu,t]\propto n^{(1+\hat\alpha)/2}\, R_\infty^2\, 
\gamma^{3\hat\alpha-1}\,
\delta^{3+\hat\alpha}\, A(\nu,t)\, \nu^{-\hat\alpha}\, ,
\label{afterglow}
\end{equation}
with $\hat\alpha$ changing from $\sim 0.5$ to $\sim 1.1$
as the emitted frequency\footnote{In the CB model, the spectral 
evolution as $\hat\alpha$ changes from $\sim\! 0.5$ to $\sim\! 1.1$
is interpolated by $(\nu/\nu_b(t))^{-0.5}/
\sqrt{1+[\nu/\nu_b(t)]^{1.1}}$ (Dado et al.~2003a).} crosses the 
{\it ``injection bend'',}
\begin{equation}
 \nu_b(t) \simeq 1.87\times 10^3\, [\gamma(t)]^3\,
\left[{n_p\over 10^{-3}\;cm^3}\right]^{1/2}\, {\rm Hz}\, .
\label{nubend}
\end{equation}
By the time $\gamma(t)=\gamma_0/2$, $F_{_{CB}}[\nu,t]$ is
orders of magnitude smaller than $F_{_{CB}}[\nu,0]$.

The attenuation $A(\nu,t)$ is a product of the attenuation in the host
galaxy, in the intergalactic medium, and in our Galaxy. The attenuation in
our galaxy in the direction of the GRB or XRF is usually estimated from the
Galactic maps of selective extinction, $E(B-V)$, of Schlegel, Finkbeiner \&
Davis (1998), using the extinction functions of Cardelli et al. (1986). 
The extinction in the host galaxy and the intergalactic medium,
$\rm A(\nu,t)$
in Eq.~(\ref{bw}), can be estimated from the difference between the
observed spectral index {\it at very early time when the CBs are still
near the SN} and that expected in the absence of extinction. Indeed, the
CB model predicts ---and the data confirm with precision--- the gradual
evolution of the effective optical spectral index towards the constant
value $\approx -1.1$ observed in all ``late'' AGs (Dado et al.
2002a, 2003a). The ``late'' index is independent of the attenuation in the
host galaxy, since at $\rm t>1$ (observer's) days after the explosion, the
CBs are typically already moving in the low-column-density, 
optically-transparent halo of the host galaxy.

The comparison of the predictions of Eq.~(\ref{afterglow}) with the
observations of optical and X-ray AG light curves and of broad-band
spectra is discussed in Dado et al.~2002a and 2003a, respectively.
The results ---for {\it all} GRBs of known redshift--- are very satisfactory
and involve a total of only five parameters (two of which, $z$ and
$\theta$ are not ``intrinsic'' to the model). 

The initial value of $\delta$ of XRFs is, on the average, smaller
by a factor $\geq 6.25$ than its mean value in GRBs. 
According to Eq.~(\ref{afterglow}), this implies that the early-time optical 
AG of XRFs should be flatter than that of long-duration GRBs, dimmer by
a factor of a few hundred, and stretched in time by a factor of a few.
At late time, when $[\gamma(t)\theta]^2\ll 1$, Eq.~(\ref{Doppler}) 
implies that $\delta(t)\approx 2\, \gamma(t)$,
while Eq.~(\ref{cubic}) implies that when $3\, [\gamma(t)\theta]^2\ll 
1$, $\gamma(t)$ approaches its asymptotic behaviour,  
$\gamma(t)\approx [2\,c\, t/ 3\, (1+z)\, x_\infty]^{-1/3}$.
Consequently, the late time 
AGs of GRBs and XRFs are similar and have the 
asymptotic behaviour $F_\nu\sim \nu^{-1.1\pm 0.1}\, t^{-2.13\pm 0.1}$
(Dado et al.~2002a). Because of the initially small Doppler factor,
the optical AGs of XRFs may already be above the injection bend 
at early time, yielding a $\sim \nu^{-1.1}$  optical spectrum
(before extinction).  

The thresholds of GRB of XRF detectors and the strong decline of the peak
luminosity with viewing angle imply that most observed XRFs 
should have relatively low redshifts. This may partly compensate for the initial 
low flux of their AGs and, combined with their relatively slow decline,
make it easier to follow their late-time AGs, particularly at radio frequencies.
Because of their relative proximity, we expect the AGs of the 
relatively-nearby XRFs to include a detectable
SN1998bw-like contribution, if  not overshined by the CBs' AG and/or by the
host galaxy. Such a SN contribution would add evidence to our claim
that XRFs and GRBs are one and the same. A signature even more
specific to the CB model would be an observed
superluminal motion of their CBs (Dar \& De R\'ujula 2000a).

So far, only the light curves of the AGs of XRFs 020903 and 030723 have
been measured and reported in sufficient detail to allow comparisons with
the CB model. We describe them next.

\section{XRF 020903}

This XRF was detected and localized with HETE II (Ricker et al.~2002).
Its AG was discovered and followed by Soderberg et al.~(2002b). Subsequent
photometric observations were reported by Covino et al.~(2002) and
Gorosabel et al.~(2002). Spectroscopic observations by Hamuy and Shectman
obtained with the Magellan 6.5m telescope 25 days after burst revealed
narrow emission lines from an underlying host galaxy at a mean redshift,
$z=0.25$ (Soderberg 2002). Subtraction of the emission lines revealed a
continuum that is consistent with a SN akin to 1998bw at $z=0.25$
(Soderberg: http://www.astro.caltech.edu/~ams/XrF.html).

\begin{figure}[t]
\vspace{-.5cm}
\hskip 2truecm
\hspace*{-2.1cm} 
\epsfig{file=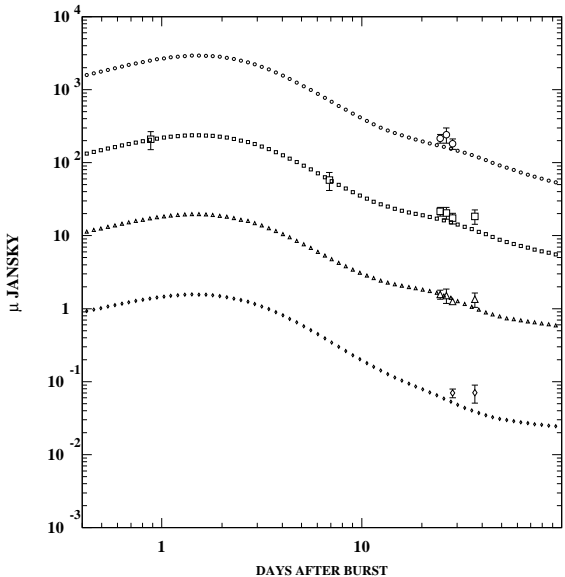, width=8.5cm} \\
\vspace{-0.5cm}
\caption{CB-model  fit to the
measured (top to bottom) I, R, V, and B-band AGs of XRF 020903,
multiplied by 10, 1, 1/10, 1/100, respectively.  
The theoretical contribution  from a SN1998bw-like supernova
at $z=0.25$
was corrected for the known extinction in the Galaxy in the direction of
XRF 020903.  Extinction in the host galaxy was neglected due to
lack of information on the early time AG.
A contribution of the host galaxy was included assuming a spectrum
of a typical star-burst galaxy.
The SN contribution is not directly visible in the light curves.
}
\label{AGBB02903}
\end{figure}

\begin{figure}[t]
\vspace{-.5cm}
\hskip 2truecm
\hspace*{-2.1cm}
\epsfig{file=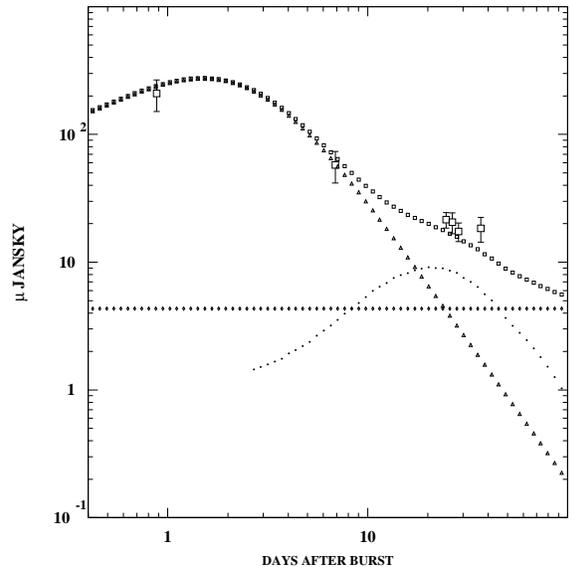, width=8.5cm} \\
\vspace{-0.5cm}
\caption{The R-band AG of XRF 020903. 
Triangles: $F_{_{CBs}}$, from the fit 
to the broad-band AG. Dots: $F_{_{SN}}$, a 1998bw-like SN
at $z=0.25$, corrected for Galactic extinction in the 
direction of
XRF 020903 but not for  extinction in the host galaxy.
Diamonds: the host galaxy, assumed to have 
the typical spectrum of a star-burst galaxy. Squares: total AG.
The SN contribution is not clearly visible in the light curve
though it dominates, in the CB model, the late time AG.}
\label{AGRB02903}
\end{figure}

The lightcurve of XRF 020903 showed a double peak structure (Sakamoto et
al.~2003 to be published), which implies, in the CB model, that it was
dominated by two CBs. Due to sparse observational data, a CB model fit
with two independent CBs does not constrain enough the fitted parameters.
Therefore, we have fitted the observations assuming a single CB, which
should be a good approximation if the two CBs had similar properties.  
Our CB-model fit to the broadband optical AG of this XRF is shown in
Fig.~(\ref{AGBB02903}) and the separate contributions of the CB, the host
galaxy and a 1998bw-like SN at $z=0.25$ are shown in
Fig.~(\ref{AGRB02903}).  Extinction in our Galaxy in the direction of XRF
020903 was included, using $E(B-V)=0.032$ Schlegel et al.~(1998), and the
phenomenological relations of Cardelli et al.~(1998), which result in the
attenuation coefficients, $A(I)=0.89$, $A(R)=0.83$, $A(V)=0.78$, and
$A(B)=0.72$. Because of lack of spectral information on the early AG, the
extinction of the SN in the host galaxy was taken to be equal to the
typical extinction in our Galaxy, i.e.~$E(B-V)=0.06$ with the extinction
relations of Cardelli et al.~(1998). The host galaxy was assumed to have
the spectrum of a template star burst galaxy, as measured e.g. by
Gorosabel et al.~(2003) for GRB 000210, with a spectral break around
$4000\, (1+z)$ \AA, and $F_\nu$ approximately constant 
at longer wavelengths.  The
fitted values of the CB model parameters are, $\gamma=1693$, $\theta=
3.66$ mrad, $n_p=3.5\times 10^{-2}$ cm$^{-3}$ and $x_{\infty}=87 $ kpc.
These parameters are all in the corresponding ranges characteristic of
GRBs, but for $\theta$ which, as expected, is larger than for all of the 32
GRBs we have analyzed, but GRB 980425.

The fact that GRB 980425 ---which, as we discussed, is marginally an
XRF--- had an even larger observer's angle than XRF 020903 is what 
explains the greater luminosity of the radio AG of the XRF. The 
direct comparison of optical AGs is not an issue because that of
XRF 020903 is due to the CBs, while the very large observation
angle of GRB 980425 implied that its optical AG was dominated 
by the SN (Dar and De Rujula 2000).

 
The poorly measured light curve of the AG of XRF 020903 does not provide a
convincing evidence for an underlying SN in XRF 020903, which dominates
its late time behaviour in the CB model, as can be seen from
Fig.~(\ref{AGRB02903}). However, precise new measurements of the spectrum
of the host galaxy are highly desirable. When subtracted from the
spectrum of the AG measured by Hamuy and Shectman on day 25 after the
burst, they may provide the decisive spectroscopic evidence for an
underlying SN akin to 1998bw in the AG of XRF 020903 (Soderberg et al.  
2002).


The equivalent isotropic $\gamma$-ray energy of XRF 020903 was estimated
to be $E_{iso}^\gamma\sim 1-2\times 10^{50}$ erg (Soderberg:
http://www.astro.caltech.edu/~ams/XrF.html).  The quoted CB-model fitted
parameters result in $\delta(t=0)\approx 87$. The
light curve of XRF 020903 appears to be dominated by two CBs (Sakamoto
et al.~2003, to be published). Thus, the
CB-model prediction for its equivalent $\gamma$-ray energy is,
$E_{iso}^\gamma=2\,\delta^3\,{\cal{E}}_{_{CB}}'
\approx 2.1\times 10^{50}$ erg, in good agreement with its
observationally-inferred value.

\section{XRF 030723}

\begin{figure}[t]
\vspace{-4.0cm}
\hskip 2truecm
\hspace*{-2.1cm}
\epsfig{file=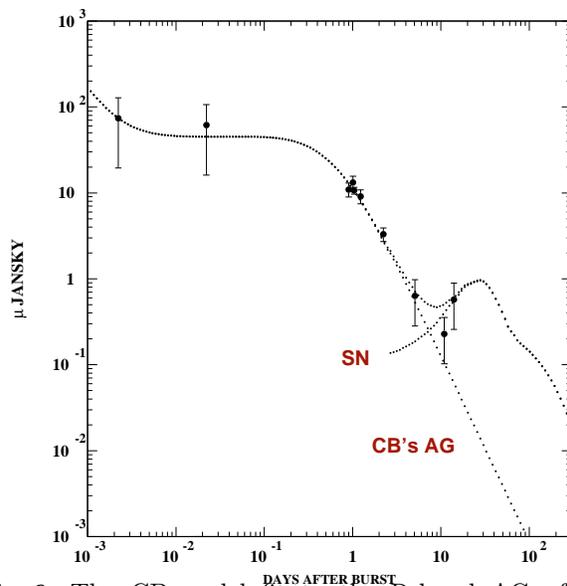, width=9cm}
\vspace{-1.5cm}
\caption{The CB-model fit to the  R-band  AG of XRF 030723 (Prigozhin     
et al.~2003;
Fox et al.~2003b; Dullighan et al.~2003a,b; Smith et al.~2003; Bond et 
al.~2003; Fynbo et al. 2003).
The first two data points have been deduced from the unfiltered measurements 
of  Smith et al.~(2003), assuming an early  $F_\nu\sim \nu^{-0.5}$ 
(Dado et al.~2003a). All errors were  multiplied
by a factor 2 to account for cross-calibration uncertainties.
The  observed ``rebrightening'' 14 days after the XRF
may have been due to the contribution of a SN
(Fynbo et al.~2003). In the CB-model's prediction,
the redshift of the 1998bw-like SN  has been
adjusted to the normalization of the late-time points;
without extinction corrections in the host galaxy or ours,
it corresponds to a redshift of $z\sim 0.75$.}
\label{figXRF}
\end{figure}

This XRF  was detected and localized with HETE II
(Prigozhin et al.~2003). Its AG was discovered by Fox et al.~(2003b). Later
measurements were reported by Dullighan et al.~(2003a,b), Smith et
al.~(2003), Bond et al.~(2003) and Fynbo et al.~(2003) who reported a 
``rebrightening" in the optical AG 14 days after the XRF, which may be due to
the contribution of a SN.  This rebrightening can be seen in
Fig.~(\ref{figXRF}), in which we also present the CB model fit to the AG. The
normalization of the 1998bw-like SN contribution has been adjusted:
without extinction corrections in the host galaxy or ours, it corresponds
to a redshift of $z\sim 0.75$ (the redshift of this XRF has not been
measured so far). The fitted values of the CB-model parameters
are $\gamma=776$, $\theta= 2.79$ mrad, $n_p=0.121$ cm$^{-3}$ and
$x_{\infty}= 29.5$ kpc. Once again, only $\theta$ is outside the
characteristic range of GRB parameters in the CB model.
 Late-time measurements of the spectrum of the host
galaxy and its redshift, as well as colour photometry and precise
spectroscopy of the late-time AG are needed to verify
whether or not the rebrightening is due to a SN akin to SN1998bw at the XRF
position.

\section{Conclusions} 

In Dar and De R\'ujula (2000) we argued that long-duration GRBs may
all be associated with core-collapse SNe akin to SN1998bw
when viewed near axis, and that GRB 980425 (nearly an XRF)
was a normal GRB viewed at a
much larger angle than ordinary GRBs. In a series of papers 
(Dado et al.~2002a,b,c, 2003a,b,c) 
we presented supporting evidence for the GRB/SN
association from a CB-model analysis of the AGs of all GRBs of
known redshift. The recent spectroscopic evidence for a GRB/SN association
in the AGs of GRBs 030329 (Stanek et al.~2003; Hjorth et al.~2003;
Matheson et al.~2003) and GRB 021211 (Della Valle et al.~2003) is perhaps
the most convincing evidence. 

In Dar \& De R\'ujula (2003) we argued that short-duration GRBs may be
associated with SNe of Type Ia, and that XRFs are ordinary GRBs, which ---like
GRB 980425--- are viewed from much larger angles. In this paper we have
demonstrated that all the currently-known properties of XRFs, including
the XRF/SN association, support this interpretation. 
If our contentions are correct, it follows that short XRFs
--unobserved to date-- should have properties that scale
relative to those of short GRBs with the same scaling laws
which relate XRFs to long GRBs. The association of short
XRFs and GRBs with SNe ---allegedly of Type Ia
and akin to SN1997cy--- is, so far, purely conjectural
(Dar \& De R\'ujula, 2003).

At the moment, the best proof of our alleged XRF/SN association may
be provided by obtaining accurate spectra of the host galaxies of XRFs
020903 and 030723, now that the putative SN and the AG have faded away. 
These spectra may be subtracted, respectively, from the 
spectrum of XRF 020903 taken
with the Magellan 6.5m telescope on September 28.1 UT 2002 (Hamuy and
Shectman 2002 to be published) near the expected SN maximum, and from the
spectrum of the AG of XRF 030723, which was measured with the VLT around
the SN maximum (Fynbo et al.~to be published). After subtraction
of the galaxies' contributions, the spectra may expose the underlying SNe.

\begin{acknowledgements} 

This research was supported in part by the Helen
Asher Space Research Fund at the Technion. Observational informations
communicated to us by J. Fynbo, M. Hamuy and A. Soderberg prior to their
publication are gratefully acknowledged.  One of us, Arnon Dar, is
grateful for the hospitality extended to him at the CERN Theory Division.

\end{acknowledgements}

\noindent
{\bf Note added}:

\noindent
After this article was posted in the astro-ph archives (Dado, Dar \& De
R\'ujula, astro-ph/0309294), Lamb, Donaghy \& Grazini reported there
(astro-ph/0309456) the peak energy, $E_p$, of 14 XRFs localized by HETE
II.  Out of these 14 cases, 6 have $E_p$ values smaller than those
observed by BeppoSAX and BATSE by about an order of magnitude. Their
fluence and isotropic equivalent energies are much larger than those
implied by Eqs.~(\ref{epflu}) and (\ref{epeiso}).  We contend that the
correct $E_p$ values for these XRFs are larger, and are similar to those
observed by BeppoSax (Heise et al.~2001) and BATSE (Kippen et al.~2002).
The very low photon fluences detected in these XRFs by HETE II above 25
keV ---as well as the unknown column density of absorbing material along
the line of sight--- must have prevented a correct determination of the
true values of $E_p$. Moreover, Lamb et al.~(2003) did not include GRB
980425 in their plot (Fig. 1.2) of $E_p$ as a function of
${\cal{E}}^{iso}$. The result for this GRB badly violates their advocated
relation, $E_p\propto [{\cal{E}}^{iso}]^{1/2}$, but is consistent with the
CB model relation: $E_p\propto [{\cal{E}}^{iso}]^{1/3}$.

{}


\begin{thebibliography}{}

\bibitem{}
Amati, L., et al. 2002, A\&A, 390, 81
\bibitem{}
Band, D., et al. 1993, ApJ, 413, 281
\bibitem{}
Barraud, C., et al. 2003, A\&A,  400, 1021
\bibitem{}
Berger, E., Kulkarni, S. \& Frail, D. A. 2003, ApJ. 590, 379
\bibitem{}
Bhat, P. N., et al. 1994, ApJ, 426, 604
\bibitem{}
Bloom, J. S., et al. 2003a, astro-ph/0302210  
\bibitem{}
Bloom, J. S., et al. 2003b, astro-ph/0303514 
\bibitem{}
Bond, H. E., et al. 2003, Nature, 422, 425
\bibitem{}
Cardelli, J, A., Clayton, G. C., Mathis, J. S. 1988, ApJ, 329, L33
\bibitem{}
Covino, S., et al. 2002, GCN Circ. 1563 
\bibitem{}
Dado, S.,  Dar, A. \& De R\'ujula, A. 2002a, A\&A, 388, 1079
\bibitem{}
Dado, S.,  Dar, A. \& De R\'ujula, A. 2002b, ApJ, 572, L143
\bibitem{}
Dado, S.,  Dar, A. \& De R\'ujula, A. 2002c, A\&A, 393, L25
\bibitem{}
Dado, S.,  Dar, A. \& De R\'ujula, A. 2003a, A\&A, 401, 243
\bibitem{}
Dado, S.,  Dar, A. \& De R\'ujula, A. 2003b, ApJ, 593, 961 
\bibitem{}
Dado, S., Dar, A. \& De R\'ujula, A. 2003c, ApJ, 594, L89
\bibitem{} 
Dado, S., Dar A. \& De R\'ujula A. 2003d, Phys. Lett. B562, 161
\bibitem{}
Dar, A. 1998, ApJ, 500, L93
\bibitem{}
Dar, A. 1999 A\&AS  138(3), 505
\bibitem{}
Dar, A. 2003, astro-ph/0301389
\bibitem{}
Dar A., De R\'ujula A., 2000, astro-ph/0008474
\bibitem{}
Dar A., De R\'ujula A., 2003, astro-ph/0308248
\bibitem{}
Dar, A. \& Plaga, R. 1999, A\&A, 349, 259
\bibitem{}
Della Valle, M., et al. 2003, A\&A, 406, L33
\bibitem{}
De R\'ujula, A. 1987,  Phys. Lett., 193, 514
\bibitem{}
De R\'ujula, A. 2002, astro-ph/0207033
\bibitem{}
De R\'ujula, A. 2003, hep-ph/0306140
\bibitem{}
Dullighan, A.,  et al. 2003a, GCN Circ. 2326
\bibitem{}
Dullighan, A.,  et al. 2003b, GCN Circ. 2336
\bibitem{}
Fenimore, E. E., in 't Zand, J. J. M., Norris, J. P., Bonnell, J. T.,
\& Nemiroff, R. J. 1995, ApJ, 448, L101
\bibitem{}
Frail, D. A., et al. 2001, ApJ, 562, L55
\bibitem{}
Frederiksen, J. T., Hededal, C. B., Haugbolle, T. \& Nordlund, A. 2003,
astro-ph/0303360
\bibitem{}
Fruchter, A., et al. 2002, GCN Circ.  1268 
\bibitem{}
Fynbo, J. P. U., et al. 2003, GCN Circ. 2345
\bibitem{}
Galama T.J., et al., 1998, Nature 395, 670
\bibitem{}
Gandolfi, G., et al. 2001, GCN Circ. 1118 
\bibitem{}
Garnavich, P. M., et al. 2003b, IAU Circ. 8108
\bibitem{}
Golenetskii, S. V., Mazets, E. P., Aptekar, R. L. \& Ilinskii, V. N.
1983, Nature, 306, 451
\bibitem{}
Gorosabel, J., et al. 2002, GCN Circ. 1631
\bibitem{}
Gorosabel, J., et al. 2003,  A\&A, 400, 127
\bibitem{}
Granot, J., Panaitescu, A., Kumar, P. \& Woosley, S. E. 2002,  2002, ApJ, 
570, L61
\bibitem{}
Hjorth, J.,  et al. 2003, Nature, 423, 847
\bibitem{}
Harrison F. A., et al. 2001, GCN Circ. 1143  
\bibitem{}
Heise, J., in't Zand, J. Kippen,, R. M., \& Woods, P. M. 2001,
Proc. 2nd Rome Workshop: GRBs in the Afterglow Era (oct,. 2000),
ed. E. Costa, F. Frontera \& J. Hjorth (springer Berlin) 
\bibitem{}
Iwamoto, K., et al. 1998, Nature, 395, 672  
\bibitem{}
Jin, Z. P. \& Wei, D. M. 2003, astro-ph/0308061
\bibitem{}
Kippen, R. M., et al. 2002, astro-ph/0203114
\bibitem{}
Lazzati, D., Rossi, E.,  Ghisellini, G. \&  Rees, M. J. 
2003, astro-ph/0309038 
\bibitem{}
Lee, A., Bloom, E. D. \& Petrosian, V., 2000, ApJS, 131, 1
\bibitem{}
McBreen,  S., Quilligan, F., McBreen, B.,
Hanlon, L.,  \& Watson, D.  2002, astro-ph/0206294
\bibitem{}
Matheson, T., et al. 2003, astro-ph/0307435
\bibitem{}
Norris, J. P., Nemiroff, R. J., Bonnell, J. T., Scargle, J. D.,
Kouveliotou, C., Paciesas, W. S., Meegan, C. A. \& Fishman, G. J. 1996,
ApJ, 459, 393
\bibitem{}
Mirabel, I. F. \& Rodriguez, L. F. 1999, ARA\&A, 37, 409
\bibitem{}
Panaitescu, A. \&  Kumar, P. 2001, ApJ, 560L, 49
\bibitem{}
Panaitescu, A. \&  Kumar, P. 2003, ApJ, 592, 390
\bibitem{}
Piran,  T.  2000, Phys. Rep. 333, 529
\bibitem{}
Plaga, R. 2001, A\&A, 370, 351
\bibitem{}
Preece, R. D., Briggs, M. S., Mallozzi, R. S.,
Pendleton, G. N., Paciesas, W. S. \& Band, D. L. 2000, ApJS, 126, 19
\bibitem{}
Prigozhin, G., et al. 2003, GCN Circ. 2313  
\bibitem{}
Quilligan, F., McBreen, B., Hanlon, L., McBreen, S.,
Hurley, K. J. \& Watson, D.  2002, A\&A, 385, 377
\bibitem{}
Ramirez-Ruiz, E. \& Fenimore, E. E. 2000, ApJ, 539, 12
\bibitem{}
Rhoads, J. E. 1997, ApJ, 487, L1
\bibitem{}
Rhoads, J. E. 1999, ApJ, 525, 737
\bibitem{}
Ricker, G., et al. 2002, GCN Circ. 1530  
\bibitem{}
Rodriguez, L. F. \& Mirabel, I. F. 1999, ApJ, 511, 398
\bibitem{}
Rossi, E., Lazzati, D. \& Rees, M. J. 2002, MNRAS, 332, 945
\bibitem{}
Salmonson, J. D. 2003, ApJ, 592, 1002
\bibitem{}
Sari, R., Piran, T. \& Narayan, R. 1998, ApJ, 497, L17
\bibitem{}
Sari, R., Piran, T. \& Halpern, J. P., 1999, 519, L17
\bibitem{}  
Schlegel D. J., Finkbeiner D. P. \&  Davis M., 1998, ApJ 500, 525
\bibitem{}
Shaviv, N. J. \& Dar, A. 1995, ApJ, 447, 863
\bibitem{}
Smith, D. A.,  et al. 2003, GCN Circ. 2338
\bibitem{}
Soderberg, A. M., et al.  2002, GCN Circ. 1554  
\bibitem{}
Stanek, K. Z., et al. 2003, ApJ,  591, L17
\bibitem{}
Taylor, G. B., et al. 2001, GCN Circ. 1136  
\bibitem{}
Waxman, E. 2003b, Nature, 423, 388
\bibitem{}
Woosley, S. E. \& MacFadyen, A. I. 1999, A\&AS 138, 499
\bibitem{}
Wu, B. \& Fenimore, E. E. 2000, ApJ, 535, L29
\bibitem{}
Yamazaki, R.,  Ioka, K. \& Nakamura, T. 2002, ApJ, 571, L31  
\bibitem{}
Yamazaki, R. Yonetoku, D. \& Nakamura, T. 2003a, ApJL, 594, L79
\bibitem{}
Yamazaki, R.,  Ioka, K. \& Nakamura, T. 2003b, ApJ, 593, 941  
\bibitem{}
in 't Zand, J., et al. 2002, GCN Circ. 1383
\bibitem{}
Zhang, B. \&  M\'esz\'aros, P. 2002, ApJ, 566, 712



\end{thebibliography}
\end{document}